\documentclass[%
 reprint,
nofootinbib,
nobibnotes,
 amsmath,amssymb,
 aps,
]{revtex4-2}
\usepackage{adjustbox}
\usepackage{graphicx}
\usepackage{dcolumn}
\usepackage{bm}
\usepackage{hyperref}
\usepackage{subcaption}

\usepackage{slashed}
\usepackage{float}
\usepackage{tabularx}
\usepackage{array}
\usepackage{subcaption}

\usepackage{ragged2e} 

\makeatletter
\renewcommand{\@makecaption}[2]{%
  \vskip\abovecaptionskip
  \sbox\@tempboxa{#1: #2}%
  \ifdim \wd\@tempboxa >\hsize
    \begin{minipage}{\linewidth}%
      \justifying\normalfont\normalsize
      \noindent\textbf{#1:} #2\par 
    \end{minipage}%
  \else
    \hbox to\hsize{\hfil \normalfont\normalsize \textbf{#1:} #2\hfil}%
  \fi
  \vskip\belowcaptionskip}
\makeatother

\begin{document}


\title{Inverse magnetic catalysis in the linear sigma model: a beyond mean field approach}%

\author{Gabriela Fernández}
 \affiliation{Departamento de F\'isica, Universidad Aut\'onoma Metropolitana-Iztapalapa, Avenida San Rafael Atlixco 186, Ciudad de México 09340, Mexico.}
\author{Ana Mizher}%
\affiliation{Instituto de Física - Universidade de São Paulo, Rua do Matão 1371, CEP 05508-090, São Paulo - Brasil.}
\author{L. A. Hernández}
\affiliation{Departamento de F\'isica, Universidad Aut\'onoma Metropolitana-Iztapalapa, Avenida San Rafael Atlixco 186, Ciudad de México 09340, Mexico.}%


\begin{abstract}
We explore the restoration of chiral symmetry in the linear sigma model coupled to quarks under the influence of strong magnetic fields and finite temperature, incorporating screening effects through ring diagrams. While previous studies using tree-level thermal masses lead to magnetic catalysis across all temperature ranges, in tension with lattice QCD results, we go beyond this limitation by computing the bosonic masses self-consistently within the lowest Landau level (LLL) approximation. The self-consistent approach modifies the effective potential and allows us to accurately track the thermal evolution of the order parameter. Our results reveal the emergence of a critical end point (CEP) in the $T-|eB|$ phase diagram and, notably, exhibit inverse magnetic catalysis (IMC) behavior: the (pseudo)critical temperature decreases with increasing magnetic field strength. This is in contrast to the magnetic catalysis behavior found when non-self-consistent masses are used. To the best of our knowledge, this is the first time that self-consistent boson masses have been implemented in this context, offering a new framework for exploring the QCD phase diagram using effective models.
\end{abstract}

\maketitle


\section{\label{sec1} Introduction}

The study of strongly interacting matter under extreme conditions in presence of magnetic fields has recently become one of the central topics in the high-energy physics community~\cite{Fukushima:2008xe,Skokov:2009qp,Miransky:2015ava,Adhikari:2024bfa,Hattori:2023egw}. In particular, significant attention has been devoted to understanding the effects of magnetic fields on such matter, both in the hadronic~\cite{Andreichikov:2013zba,Machado:2013yaa,Mueller:2014tea,Liu:2014uwa,Taya:2014nha,Simonov:2015xta,Gubler:2015qok,Hattori:2015aki,Yoshida:2016xgm,Zhang:2016qrl,Ayala:2016bbi,Ghosh:2016evc,Bali:2017ian,Ghosh:2017rjo,Coppola:2018vkw,Liu:2018zag,Aguirre:2018fbo,Ayala:2018zat,Avancini:2018svs,Carlomagno:2022inu,Ayala:2023llp} and deconfined phases~\cite{Zayakin:2008cy,Frasca:2011zn,Ayala:2014uua,Hattori:2017xoo,Ayala:2018wux,Ayala:2019akk,Ayala:2020wzl,Ballon-Bayona:2020xtf,Fernandez:2024tuk}. Moreover, these magnetic effects have been analyzed in the context of the Quantum Chromodynamics (QCD) phase transition, understood as a transition between a hadron gas and the quark-gluon plasma~\cite{DElia:2010abb,Mizher:2010zb,Bali:2011qj,Fraga:2012ev,Bali:2012zg,Fukushima:2012kc,Bali:2014kia,Ayala:2015bgv,Endrodi:2015oba,Pagura:2016pwr,Bandyopadhyay:2020zte,DElia:2021yvk,Ayala:2021nhx,Andersen:2021lnk}. Physical systems where such extreme conditions and thus non-negligible magnetic effects are relevant include the early universe, the core of neutron stars, and non-central relativistic heavy-ion collisions.

A turning point in the study of strongly interacting matter in the presence of magnetic fields was the identification of the phenomenon known as Magnetic Catalysis (MC)\cite{Bali:2012zg,Gusynin:1994re,Gusynin:1995nb}. This effect manifests in the QCD vacuum as an enhancement of the light-quark condensate with increasing magnetic field strength, indicating that the magnetic field catalyzes chiral symmetry breaking in QCD. Subsequent studies extended this analysis to finite temperature, exploring magnetic effects on strongly interacting matter in a thermal medium. Around the QCD phase transition at zero baryon chemical potential, it was found that the pseudocritical temperature decreases as the magnetic field strength increases. Simultaneously, under the same thermodynamic conditions, the light-quark condensate also decreases with the magnetic field. This latter effect is referred to as Inverse Magnetic Catalysis (IMC), and it was first reported by Lattice QCD (LQCD) simulations\cite{Bali:2011qj,Bali:2012zg,Bali:2014kia}. The discovery of both MC and IMC triggered a broad line of research aimed at understanding their underlying mechanisms.

As a result of more systematic studies of IMC, efforts were directed toward constructing the QCD phase diagram in the presence of magnetic fields. Initial investigations focused on identifying how the transition line is modified and whether the nature of the phase transition changes in the phase diagram defined in the $T-\mu_B$ plane~\cite{Costa:2013zca,Ayala:2015lta,Ferreira:2017wtx,Ayala:2021nhx,Braguta:2019yci,Moreira:2021ety}. In parallel, an alternative version of the phase diagram was developed, defined in the $T-|eB|$ plane~\cite{Bali:2011qj,Andersen:2014xxa,Endrodi:2015oba,Gatto:2010pt,DElia:2021yvk,Bandyopadhyay:2019pml}. The latter is the phase diagram of interest in this work. In Ref.~\cite{DElia:2021yvk}, it was observed that the pseudocritical temperature $T_c$ decreases as the magnetic field strength increases, consistent with the IMC phenomenon. However, an additional and noteworthy result was reported: the appearance of a critical point at very large values of $|eB|$. Recall that at zero baryon chemical potential, LQCD results indicate a crossover\cite{Aoki:2006we,Borsanyi:2020fev}. Therefore, the presence of a strong magnetic field induces a change in the nature of the QCD phase transition. This novel feature complements the IMC and further enriches the structure of the QCD phase diagram. 

In order to complement the study of the QCD phase diagram in the $T-|eB|$ plane performed by LQCD, various effective models, such as the Nambu–Jona-Lasinio (NJL) model, the linear sigma model (LSM), and their extensions, have been employed in attempts to reproduce the LQCD results and to understand the origin of the observed phenomena. However, to date, there have been no reports of effective model calculations (either beyond or within mean-field approximations and without introducing effective couplings) that simultaneously reproduce both inverse magnetic catalysis and the emergence of a critical point. This highlights the importance of considering a running interaction strength that depends on temperature and magnetic field. While some studies have managed to qualitatively reproduce LQCD results by introducing effective couplings whose form is externally prescribed rather than derived from the model’s internal dynamics, this approach sacrifices theoretical consistency for empirical agreement. The widespread use of these externally guided interactions to match LQCD findings underscores the need for an alternative strategy. The present work aims to address this by offering a framework that captures the essential physics without relying on parametrizations imposed from outside the model. To this end, we implement a self-consistent computation of the quasi-particle masses, allowing us to account for the full effects of temperature and magnetic field on the dynamics of the system while remaining entirely within the structure of the effective model.

In this work, we aim to construct the QCD phase diagram in the $T-|eB|$ plane up to ring-diagrams order, computing the masses of the fields in a self-consistent manner within the framework of the linear sigma model coupled to quarks (LSMq). To achieve this goal, we adopt a two-step strategy. In the first part of our analysis, we construct the effective potential including tree-level field-dependent masses, and evaluate its behavior under the influence of finite temperature and an external magnetic field. This preliminary approach allows us to identify general features of the phase structure and to assess the emergence of a critical endpoint. However, as will be shown, this treatment leads to magnetic catalysis only, that is, 
an enhancement of the chiral symmetry breaking for all ranges of temperature which contradicts LQCD results~\cite{Endrodi:2015oba,DElia:2021yvk}. In the second part of the work, we implement a self-consistent determination of the boson masses by computing the curvature of the effective potential. This improved approach accounts for collective effects and introduces dynamical thermal and magnetic corrections to the masses, enabling a more accurate description of the system's response to external conditions. As we demonstrate, this procedure successfully recovers both the phenomenon of inverse magnetic catalysis and the existence of a critical point, in qualitative agreement with LQCD findings~\cite{Endrodi:2015oba,DElia:2021yvk}.

The paper is organized as follows. In Section~\ref{sec2}, we introduce the LSMq and discuss its relevant features for this study. We present the Lagrangian density and describe the spontaneous breaking of chiral symmetry. In Section~\ref{sec3}, we compute all contributions to the effective potential (or free energy) up to ring-diagrams order using the Matsubara formalism. Magnetic field effects are incorporated through the propagators of charged particles, and we adopt the lowest Landau level (LLL) approximation, under the assumption that the magnetic field is the dominant energy scale. We also analyze the validity of these approximations. In Section~\ref{sec4}, we describe the self-consistent computation of the masses of the fields. This procedure constitutes a novel contribution to the analysis of chiral symmetry restoration in a magnetic background and is expected to significantly improve the predictive power of the model (for a similar treatment in the absence of magnetic field see \cite{Scavenius:2000qd}). Section~\ref{sec5} presents our results, including a systematic analysis aimed at mapping the effective QCD phase diagram in the $T-|eB|$ plane. Finally, in Section~\ref{sec6}, we summarize our findings and present our conclusions.

\section{\label{sec2} Linear sigma model coupled to quarks }

The LSMq is one of the most successful effective models for describing the low-energy regime of QCD. A key feature of this renormalizable model is its ability to exhibit spontaneous symmetry breaking. Its degrees of freedom consist of a mixture of scalar and pseudoscalar mesons, along with the two lightest quark flavors. The Lagrangian density of this model is given by
\begin{align}
\mathcal{L} &= \frac{1}{2}(\partial_{\mu}\sigma)^2+\frac{1}{2}(\partial_{\mu} \pi_{0})^2+D_{\mu} \pi_- D^{\mu} \pi_{+}\nonumber \\
&+\frac{a^2}{2}(\sigma^2+\pi_0^2+2 \pi_{-}\pi_{+})-\frac{\lambda}{4} \left( \sigma^2+\pi_0^2+2 \pi_{-}\pi_{+}\right) \nonumber \\
&+i\bar{\psi} \slashed{\partial}\psi - g\bar{\psi} ( \sigma+i \gamma^{5} \vec{\tau}.\vec{\pi} ) \psi,
\label{lagrangian}
\end{align}
where $\psi$ is an $SU(2)$ isospin doublet of quarks, $\sigma$ is an isospin singlet and $\vec{\pi}=(\pi_+, \pi_-, \pi_0 )$ is an isospin triplet, corresponding to the sigma meson, the two charged pions and one neutral pions, respectively. In Eq.~(\ref{lagrangian}), $\tau$ are the Pauli matrices. Also, two different couplings appear, $\lambda$ and $g$, the boson self-coupling and the fermion-boson coupling, respectively. The squared mass parameter is $a^2$. In this work, we take $a^2,\lambda,g >0$. In order to allow for a spontaneous symmetry breaking, the $\sigma$ field develops a vacuum expectation value $v$, namely
\begin{equation} 
\sigma \rightarrow \sigma + v.
\label{shift}
\end{equation} 
Thus, we rewrite the Lagrangian in Eq.~(\ref{lagrangian}) as follows
\begin{align}
    \mathcal{L}&=\frac{1}{2}\partial_{\mu}\sigma \partial^{\mu}\sigma+\frac{1}{2}\partial_{\mu}\pi_{0}\partial^{\mu}\pi_{0}+D_{\mu}\pi_{-}D^{\mu}\pi_{+}\nonumber\\
    &-\frac{1}{2}m_{\sigma}^{2}\sigma^{2}-\frac{1}{2}m_{\pi}^{2}\pi_{0}^{2}-m_{\pi}^{2}\pi_{-}\pi_{+}+i\bar{\psi}\slashed{\partial}\psi\nonumber\\
    &-m_{f}\bar{\psi}\psi+\frac{a^2}{2}v^2-\frac{\lambda}{4}v^4+\mathcal{L}_{int},
    \label{linearsigmamodelSSB}
\end{align}
where the interaction Lagrangian is defined as
\begin{align}
    \mathcal{L}_{int}&=-\frac{\lambda}{4}\sigma^{4}-\lambda v\sigma^{3}-\lambda v^{3}\sigma-\lambda\sigma^{2}\pi_{-}\pi_{+} -2\lambda v \sigma\pi_{-}\pi_{+} \nonumber \\
    &-\frac{\lambda}{2}\sigma^{2}\pi_{0}^{2}-\lambda v\sigma \pi_{0}^{2}-\lambda \pi_{-}^{2}\pi_{+}^{2}-\lambda\pi_{-}\pi_{+}\pi_{0}^{2}-\frac{\lambda}{4}\pi_{0}^{4} \nonumber \\ 
    &+a^{2}v\sigma -g\bar{\psi}\psi\sigma-ig\gamma^{5}\bar{\psi}\left(\tau_{+}\pi_{+}+\tau_{-}\pi_{-}+\tau_{3}\pi_{0}\right)\psi.
    \label{interactinglagrangian}
\end{align}

Notice from Eq.~(\ref{linearsigmamodelSSB}), there are new terms associated to the masses of the fields which are
\begin{equation}
     m_{\sigma}^{2}=3\lambda v^2-a^2, \ \ \
     m_{\pi}^{2}=\lambda v^2-a^2, \ \ \ 
     m_{f}=gv.
\label{masses}
\end{equation}
These masses are dynamically generated and depend on the vacuum expectation value, $v$, which serves as the order parameter associated with the spontaneous breaking of chiral symmetry. As a consequence of this symmetry breaking, we can parametrize the potential along the sigma-meson field direction. At the classical (or tree) level, it is written as
\begin{equation}
    V^{\text{tree}}(v)=-\frac{a^2}{2}v^2+\frac{\lambda}{4}v^4,
    \label{treelevel}
\end{equation}
whose minimum is found at
\begin{equation}
    v_0=\sqrt{\frac{a^2}{\lambda}}.
\end{equation}

Since the goal of this work is to study the effects of a magnetic field on chiral symmetry restoration and, consequently, to determine an effective QCD phase diagram, an external uniform and constant magnetic field is introduced into the model. This is achieved by incorporating a covariant derivative in the Lagrangian density for the charged fields, namely
\begin{equation}
    \partial_\mu \rightarrow D_\mu=\partial_\mu+iqA_\mu,
    \label{covariantder}
\end{equation}
where $A_\mu$ is the vector potential corresponding to an external magnetic field along the $\hat{z}$ axis, and $q$ is the charge of the field.

Since $v_0\neq 0$, we observe that chiral symmetry is spontaneously broken. To determine the conditions for chiral symmetry restoration as a function of $T$ and $|eB|$, we study the behavior of the effective potential. For this work, it includes tree-level contributions, one-loop corrections for both bosons and fermions, and the ring diagrams contribution, which accounts for plasma screening effects. In the next section, we compute each of these contributions and combine them to write the expression for the effective potential.

\section{\label{sec3} Effective potential}

Our aim is to construct the effective potential within the LSMq including thermal and magnetic corrections, and to use it as a tool to analyze the chiral phase transition in extreme conditions. We proceed in two stages. First, we compute the effective potential using field-dependent masses defined at tree level, in order to identify general features of the phase structure. As will be shown, this approach predicts the existence of a critical endpoint, but leads to magnetic catalysis, in disagreement with lattice QCD results~\cite{DElia:2021yvk}. In the second stage, we improve the analysis by computing the boson masses in a self-consistent manner from the curvature of the effective potential. This allows us to capture collective effects and to recover both inverse magnetic catalysis and the presence of a critical point, in qualitative agreement with LQCD. We now present the explicit computation of each contribution to the effective potential up to the ring-diagram level 
\begin{equation}
V^{\text{eff}}=V^{\text{tree}}+V^1_b+V^1_f+V^{\text{ring}},
    \label{Veffelements}
\end{equation}
where $V^{\text{tree}}$ is the classical potential, given by Eq.~(\ref{treelevel}). In the present case, however, we include a term that explicitly breaks chiral symmetry. Thus, the tree-level potential becomes
\begin{equation}
    V^{\text{tree}}(v)=-\frac{a^2}{2}v^2+\frac{\lambda}{4}v^4-hv,
    \label{treelevel-explicitbreaking}
\end{equation}
with $h=m_0^2v_0$, where $m_0$ is the vacuum pion mass and $v_0$ denotes the vacuum expectation value of the sigma field at tree level. The term $V^1_b$ corresponds to the one-loop bosonic contribution, $V^1_f$ to the one-loop fermionic contribution, and $V^{\text{ring}}$ accounts for the beyond-mean-field corrections associated with the ring diagrams. 

\subsection{\label{subsec3.1} Boson contribution.}
We proceed to compute the quantum fluctuations from the bosonic fields at finite temperature, order by order, using the Matsubara formalism. The one-loop bosonic contribution can be separated into two cases: the contribution from neutral bosons and the contribution from charged bosons in the presence of a magnetic field. Both cases are treated in parallel in the following discussion.

The one-loop contribution from neutral bosons is given by
\begin{equation}
    V_b^{1,0}=\frac{T}{2} \sum_n \int \frac{d^3k}{(2\pi)^3} \ln \big[G(\omega_n,\vec{k})^{-1}\big],
    \label{V10initial}
\end{equation}
where the propagator for neutral scalar fields is
\begin{equation}
    G(\omega_n,\vec{k})=-\frac{i}{\omega^2_n+\vec{k}^2+m_b^2},
    \label{scalarpropneutro}
\end{equation}
where $m_b$ represents the mass of the bosonic degrees of freedom and $\omega_n$ are the Matsubara frequencies. On the other hand, the one-loop contribution from charged bosons in the presence of a constant magnetic field is
\begin{equation}
    V_b^{1,B}=\frac{T}{2} \sum_n \int \frac{d^3 k }{(2\pi^3)} \ln  \big[G^{\text{LLL}}(\omega_n, \vec{k},|eB|)^{-1}\big],
    \label{V1Binitial}
\end{equation}
where the propagator for charged scalar fields in the lowest Landau level (LLL) approximation reads
\begin{equation}
    G^{LLL}(k)=2 i  \frac{e^{-\frac{k_{\perp}^2}{|eB|}}}{{\omega_n^2+k_{3}^2+m_b^2+|eB|}}.
    \label{scalarpropcharged}
\end{equation}
Here, $k_\perp$ and $k_3$ denote the momentum components transverse and parallel to the magnetic field, respectively. Throughout this work, we assume that the magnetic field is sufficiently strong for the LLL approximation to be valid.

To facilitate the evaluation of Eqs.~(\ref{V10initial}) and~(\ref{V1Binitial}), it is convenient to rewrite both expressions by differentiating and subsequently integrating with respect to the squared mass $m_b^2$. For neutral bosons, we obtain
\begin{equation}
    V_b^{1,0}=\frac{T}{2}\sum_n\int \frac{d^3k}{(2\pi)^3} dm_b^2 \frac{1}{\omega^2_n+\vec{k}^2+m_b^2},
\end{equation}
whereas for charged bosons, the corresponding expression is
\begin{equation}
    V_b^{1,B}=\sum_n T \int \frac{d^3k}{(2\pi)^3} dm_b^2 \frac{e^{-\frac{k_{\perp}^2}{|eB|}}}{\omega_n^2+k_3^2+m_b^2+|eB|}.
\end{equation}
Both expressions are now written in a form that allows the Matsubara frequency sums to be performed straightforwardly.

Carrying out the Matsubara frequency sums in both contributions, we find that the one-loop term for neutral bosons becomes
\begin{align}
    V_b^{1,0}=\frac{1}{8} \int \frac{d^3k}{(2\pi)^3} dm_b^2 &\frac{1}{\sqrt{\vec{k}^2+m_b^2}}\nonumber \\
    &\Bigg( 1+\frac{2}{e^{\frac{\sqrt{\vec{k}^2+m_b^2}}{T}}-1}\Bigg)
    \label{V10afterMatsubara}
\end{align}
whereas for charged bosons in the presence of a magnetic field, the corresponding expression reads
\begin{align}
    V_b^{1,B}=\frac{1}{2}\int \frac{d^3k}{(2\pi)^3} d m_b^2 &\frac{e^{-\frac{k_{\perp}^2}{|eB|}}}{\sqrt{k_3^2+m_b^2+|eB|}}\nonumber \\
    &\Bigg ( 1+\frac{2}{e^{\frac{\sqrt{k_3^2+m_b^2+|eB|}}{T}}-1}\Bigg)
    \label{V1BafterMatsubara}
\end{align}
In both cases, Eqs.~(\ref{V10afterMatsubara}) and~(\ref{V1BafterMatsubara}), it is now clear that the one-loop contributions naturally separate into two parts: a vacuum contribution, which is independent of temperature, and a matter contribution, which contains the explicit temperature dependence.

The vacuum contributions correspond to the temperature-independent terms in Eqs.~(\ref{V10afterMatsubara}) and~(\ref{V1BafterMatsubara}). These terms exhibit ultraviolet (UV) divergences arising from the integration over the momentum components. To isolate and handle these divergences, we apply dimensional regularization to the momentum integrals. Within the $\overline{MS}$ subtraction scheme, we perform the renormalization, which introduces a dependence on the renormalization scale $\mu$. For neutral bosons, the renormalized vacuum contribution to the effective potential is found to be 
\begin{equation}
    V_{b,vac}^{1,0}=-\frac{m_b^4}{64 \pi^2}\left( \frac{3}{2}+\ln\left( \frac{\mu^2}{m_b^2}\right)\right).
    \label{V1vacneutralbosons}
\end{equation}
For charged bosons, following the same regularization and renormalization procedure, the vacuum contribution reads
\begin{equation}
    V_{b,vac}^{1,B}=\frac{|eB|}{(4\pi)^2}\big(m_b^2+|eB|\big) \left(  1+ \ln \left( \frac{\mu^2}{m_b^2+|eB|}\right)\right).
    \label{V1vacchargedbosons}
\end{equation}

The matter contributions correspond to the temperature-dependent terms in Eqs.~(\ref{V10afterMatsubara}) and~(\ref{V1BafterMatsubara}). Unlike the vacuum parts, these terms are finite and do not require regularization. However, the momentum integrals cannot be computed analytically in closed form in general. Therefore, two complementary approaches are adopted to evaluate them: the high-temperature expansion and numerical integration.

For neutral bosons, the matter contribution can be treated analytically under the high-temperature approximation, leading to the well-known expression~\cite{Kapusta:2007xjq}
\begin{align}
    V_{b,HT}^{1,0}&=-\frac{T^4 \pi^2}{90}+\frac{m_b^2T^2}{24}-\frac{m_b^3 T}{12\pi}\nonumber \\
    &-\frac{m_b^4}{64\pi^2}\left(2 \gamma_{E}-\frac{3}{2} +\ln \left( \frac{m_b^2}{(4\pi T)^2}\right) \right),
    \label{V1neutralbosonsHT}
\end{align}
where $\gamma_E$ is the Euler–Mascheroni constant. Alternatively, the full matter contribution for neutral bosons can be evaluated numerically, resulting in the expression
\begin{equation}
    V_{b,N}^{1,0}=\frac{1}{2} \int \frac{d^3k}{(2\pi)^3} dm_b^2 \frac{1}{\sqrt{\vec{k}^2+m_b^2}} \frac{1}{e^{\sqrt{\vec{k}^2+m_b^2}/T}-1}.
    \label{V1neutralbosonsN}
\end{equation}
For charged bosons, the high-temperature approximation yields the expression
\begin{equation}
    V_{b,HT}^{1,B}=-\frac{T|eB|}{2\pi^2} \sqrt{m_b^2+|eB|} \sum_{n=1}^{\infty} \frac{K_1\left( n\sqrt{m_b^2+|eB|}/T\right)}{n},
    \label{V1chargedbosonsHT}
\end{equation}
where $K_1$ is the modified Bessel function of the second kind. The detailed derivation of this expression is provided in Appendix~\ref{appendix2}. The full numerical evaluation of the matter contribution from charged bosons reads
\begin{equation}
    V_{b,N}^{1,B}=\frac{T |eB|}{2\pi} \int \frac{dk_3}{2\pi} \ln \left( 1- e^{-\sqrt{k_3^2+m_b^2+|eB|}/T} \right).
    \label{V1chargedbosonsN}
\end{equation}

Thus, for both neutral and charged bosons, the matter contributions can be evaluated either through an analytic expansion valid at high temperature or by means of a full numerical computation.

Having computed the contributions from both neutral and charged bosons at one-loop order, we now have two complementary forms for the effective potential, incorporating both vacuum and matter sectors. The analytic expressions provide reliable results in the high-temperature regime, while the numerical evaluation allows access to the full thermodynamic behavior without approximations.

These results complete the construction at one-loop order of the bosonic sector contribution to the effective potential, setting the stage for the inclusion of the fermionic sector and the analysis of plasma screening effects through the ring-diagrams contribution, which we address in the following sections.

\subsection{Fermion contribution}

We now compute the one-loop fermion contribution to the effective potential. In the presence of a magnetic field the expression reads
\begin{equation}
    V_f^1=- N_c T \sum_n \int \frac{d^3k}{(2\pi)^3} \text{Tr}[S^{LLL}(\tilde{\omega}_n,\vec{k})^{-1}],
\end{equation}
where $N_c$ is the number of colors and the propagator for charged fermion fields with spin $1/2$, in the lowest Landau level approximations reads
\begin{equation}
    iS_f^{LLL}(\tilde{\omega}_n,\vec{k})=-2i e^{-\frac{k_\perp^2}{2|qB|}}\frac{i\gamma_0\tilde{\omega}_n-\gamma_3k^3+m_f}{\tilde{\omega}_n^2+k_3^2+m_f^2}\mathcal{O}^{\pm},
\end{equation}
where $m_f$ is the fermion mass, $q$ is its electric charge, $\tilde{\omega}_n=(2n+1)\pi T$ are the fermionic Matsubara frequencies, and $\mathcal{O}^{\pm}\equiv \frac{1}{2}(1\pm i \gamma^1\gamma^2)$ are the spin projectors. As we did in the bosonic case, we can rewrite the one-loop fermion contribution to the effective potencial, differentiating and subsequently integrating with respect to the squared mass $m_f^2$. Therefore, we get
\begin{equation}
    V_f^1=-4N_c \sum_n T \int \frac{d^3k}{(2\pi)^3} dm_f^2 \frac{e^{-\frac{k_\perp^2}{|qB|}}}{\tilde{\omega}_n^2+k_3^2+m_f^2}.
    \label{GeneralVf1}
\end{equation}
Performing the Matsubara sum yields
\begin{align}
    V_f^1=&-2 N_c \int \frac{d^3k}{(2\pi)^3} dm_f^2  \frac{e^{-\frac{k_\perp^2}{|qB|}}}{\sqrt{k_3^2+m_f^2}}\nonumber \\
    &\left(  1-\frac{2}{e^{\sqrt{k_3^2+m_f^2}/T}+1} \right).
    \label{Vf1afterMatsubarasum}
\end{align}
As in the bosonic case, the integrand separates naturally into vacuum and thermal contributions. The vacuum term contains an ultraviolet divergence, which we regularize using dimensional regularization and renormalize via the $\overline{MS}$ substraction scheme. The result is
\begin{equation}
    V_{f,vac}^1=-\frac{N_c |qB|}{4\pi^2}m_f^2\left(  1+\ln \left( \frac{\mu^2}{m^2_f}\right)\right),
    \label{V1vacfermion}
\end{equation}
where $\mu$ is the renormalization scale.

The thermal contribution cannot be evaluated analytically in general, and we again consider two approaches. Within the high-temperature regime, we obtain
\begin{equation}
    V_{f,HT}^1=-\frac{2N_c|qB| T }{\pi^2}m_f\sum_{n=1}^{\infty} \frac{(-1)^{n-1}}{n} K_1\left(\frac{n m_f}{T}\right),
    \label{V1fHT}
\end{equation}
where $K_1(x)$ is the modified Bessel function of the second kind. The derivation of this expression is given in Appendix~\ref{appendix3}. Alternatively, the thermal part can be evaluated numerically from
\begin{equation}
    V_{f,N}^1=\frac{N_c |qB|}{\pi}\int \frac{dk_3}{2\pi } dm_f^2 \frac{1}{\sqrt{k_3^2+m_f^2}}\frac{1}{e^{\sqrt{k_3^2+m_f^2}/T}+1}.
    \label{V1fnumeric}
\end{equation}

With the results presented so far, we have all the contributions to the effective potential at one-loop order, which corresponds to the mean-field theory approximation. While this level of approximation captures important thermal and quantum effects, it is not sufficient to fully describe the behavior of an interacting system, particularly one that exhibits collective phenomena. One way to go beyond mean-field theory is through the inclusion of ring-diagrams correction, which account for screening effects that become increasingly relevant at high temperature. These corrections are essential not only to capture collective dynamics but also to restore the analyticity of the effective potential, as they regulate infrared-sensitive terms such as those proportional to $m_b^3$ that appear in the high temperature expansion, which can otherwise become imaginary.

\subsection{Ring-diagrams contribution}

To account for screening effects, we include the contribution from ring diagrams. The starting point is
\begin{equation}
    V^{ring}=\frac{T}{2}\sum_n \int \frac{d^3k}{(2\pi)^3} \ln \left( 1+\Pi~ G \right),
    \label{Vringinitial}
\end{equation}
where $\Pi$ denotes the boson self-energy and $G$ is the propagator of the corresponding bosonic field from Eq.~(\ref{scalarpropneutro}). This expression can be recast as
\begin{equation}
    V^{ring}=\frac{T}{2} \sum_n \int \frac{d^3k}{(2\pi)^3}\left\{ \ln \left( G^{-1}+\Pi \right)-\ln G^{-1} \right\},
    \label{Vringrewritten}
\end{equation}
highlighting that ring diagrams represent a resummation of perturbative corrections to the boson propagator. In the high-temperature limit, the Matsubara zero mode ($n=0$) dominates the sum, simplifying the expression. After differentiating and integrating with respect to the squared boson mass, the ring contribution becomes
\begin{equation}
    V^{ring}=\frac{T}{2} \int \frac{d^3k}{(2\pi)^3} dm_b^2 \left(  \frac{1}{k^2+m_b^2+\Pi} -\frac{1}{k^2+m_b^2}\right),
    \label{Vringmode0}
\end{equation}
which leads to the analytic result
\begin{equation}
    V^{ring}= -\frac{T}{12\pi}(m_b^2+\Pi)^{3/2}+\frac{T m_b^3}{12\pi}.
    \label{Vringfinal}
\end{equation}
This expression explicitly requires the evaluation of the boson self-energy $\Pi$, which is computed in the next subsection.

\begin{widetext}The ring-diagrams contribution is the final term required to construct the full effective potential. Hence, the complete expression reads
\begin{align}
    V^{\text{eff}}_{HT}=&-\frac{a^2+\delta a^2}{2}v^2+\frac{\lambda+\delta\lambda}{4}v^4-h v-\frac{m_\sigma^4}{64 \pi^2}\left(2\gamma_E+\ln\left( \frac{\mu^2}{(4\pi T)^2}\right)\right)-\frac{m_\pi^4}{64 \pi^2}\left(2\gamma_E+\ln\left( \frac{\mu^2}{(4\pi T)^2}\right)\right)\nonumber \\
    &-\frac{T^4 \pi^2}{45}+\frac{m_\sigma^2T^2}{24}+\frac{m_\pi^2T^2}{24}-\frac{(m_\sigma^2+\Pi_{\sigma})^{3/2} T}{12\pi}-\frac{(m_\pi^2+\Pi_{\pi_0})^{3/2} T}{12\pi}\nonumber\\
    &+\frac{|eB|}{2\pi^2}(m_\pi^2+|eB|) \left (  1+ \ln \left ( \frac{\mu^2}{m_\pi^2+|eB|}\right )\right )-\frac{T|eB|}{\pi^2} \sqrt{m_\pi^2+|eB|} \sum_{n=1}^{\infty} \frac{1}{n}K_1\left ( \frac{n\sqrt{m_\pi^2+|eB|}}{T}\right )\nonumber\\
    &-\frac{3 |qB| }{4\pi^2} m_f^2\left ( 1+\ln\left ( \frac{\mu^2}{m_f^2}\right )\right ) -\frac{6|qB| T }{\pi^2} m_f \sum_{n=1}^{\infty} \frac{(-1)^{n-1}}{n}K_1\left (\frac{ n m_f}{T}\right ),
    \label{fulleffectivepotentialHT}
\end{align}
for the high-temperature approximation, and
\begin{align}
    V^{\text{eff}}_{N}=& -\frac{a^2+\delta a^2}{2}v^2+\frac{\lambda+\delta \lambda}{4}v^4-h v-\frac{m_\sigma^4}{64 \pi^2}\left( \frac{3}{2}+\ln\left( \frac{\mu^2}{m_\sigma^2}\right)\right)-\frac{m_\pi^4}{64 \pi^2}\left( \frac{3}{2}+\ln\left( \frac{\mu^2}{m_\pi^2}\right)\right) \nonumber \\
    &+\frac{T}{2 \pi^2}\int_0^{\infty} dk\, k^2 \ln\left(1-e^{-\sqrt{k^2+m_\sigma^2+\Pi_{\sigma}}/T}\right)+\frac{T}{2 \pi^2}\int_0^{\infty} dk\, k^2 \ln\left(1-e^{-\sqrt{k^2+ m_\pi^2+\Pi_{\pi_0}}/T}\right) \nonumber \\
    &+\frac{|eB|}{2\pi^2}(m_\pi^2+|eB|) \left (  1+ \ln \left ( \frac{\mu^2}{m_\pi^2+|eB|}\right )\right )-\frac{T|eB|}{\pi^2} \sqrt{m_\pi^2+|eB|} \sum_{n=1}^{\infty} \frac{1}{n}K_1\left ( \frac{n\sqrt{m_\pi^2+|eB|}}{T}\right )\nonumber\\
    &-\frac{3 |qB| }{4\pi^2} m_f^2\left ( 1+\ln\left ( \frac{\mu^2}{m_f^2}\right )\right ) -\frac{6|qB| T }{\pi^2} m_f \sum_{n=1}^{\infty} \frac{(-1)^{n-1}}{n}K_1\left (\frac{ n m_f}{T}\right ),
    \label{fulleffectivepotentialN}
\end{align}for the numerical evaluation. 
\end{widetext}

It is clear from both expressions that studying the features of this effective potential requires the explicit form of the boson self-energies $\Pi$ and a suitable choice of the model's free parameters: $\lambda$, $g$ and $a$. These parameters are fixed in vacuum using the following relations
\begin{equation}
    \lambda=\frac{\mathbf{m}_\sigma^2-\mathbf{m}_\pi^2}{2f_\pi^2}, \ \ \
    a=\sqrt{\lambda f_\pi^2-\mathbf{m}_\pi^2}, \ \ \
    g=\sqrt{\frac{\lambda}{2}},
    \label{fixedparameters}
\end{equation}
where $\mathbf{m}_\sigma$, $\mathbf{m}_\pi$ and $f_\pi$ denote the physical masses of the sigma meson and pion, and the pion decay constant, respectively. In Eqs.~(\ref{fulleffectivepotentialHT}) and~(\ref{fulleffectivepotentialN}), we also identify two additional constants, $\delta a^2$ and $\delta \lambda$, which are counter-terms introduced to ensure that the $T$-independent one-loop radiative corrections do not shift the minimum or the sigma meson mass from their tree-level values~\cite{Carrington:1991hz}.

Since not all the contributions have been made explicit so far, in the next section, we provide the remaining ingredients required for the full evaluation of the effective potential, namely, the explicit expressions for the boson self-energies.

\subsection{Boson self-energy}

The one-loop boson self-energies are shown diagrammatically in Fig.~\ref{diagramsselfenergy}. The corresponding expressions are given by


\begin{align}
\Pi_{\sigma}&=\frac{\lambda}{4} \left[ 12 I(m_{\sigma}^2) +4 I (m_{\pi_0}^2)+8 I (m_{\pi_{\pm}}^2)\right] +N_f N_c \Pi_f, \nonumber \\
\Pi_{\pi_0}&=\frac{\lambda}{4} \left[ 4 I (m_{\sigma}^2) +12 I (m^2_{\pi_0})+8I (m^2_{\pi_{\pm}})\right] +N_f N_c \Pi_f,\nonumber \\
\Pi_{\pi_{\pm}}&=\frac{\lambda}{4} \left[ 4 I (m^2_{\sigma}) +4 I ( m^2_{\pi_0})+16 I (m^2_{\pi_{\pm}})\right] +N_f N_c \Pi_f,
\label{self-energies}
\end{align}
where the loop integrals are defined by
\begin{equation}
    I( m_b^2)=2 \frac{d V^1}{d m_b^2},
    \label{self-energy-potentialbosons}
\end{equation}
with $b=\sigma, \ \pi_0, \ \pi_\pm$, and $V^1$ denoting the matter contribution of the one-loop effective potential. The quark-antiquark loop contribution is given by
\begin{equation}
    \Pi_f=2g^2 \frac{d V^1}{d m_f^2},
    \label{self-energy-potentialfermions}
\end{equation}

and $N_f$ and $N_c$ are the number of quarks flavors and colors, respectively. \begin{widetext}
\center{
\begin{figure}[t]
\centering
    \includegraphics[width=0.9\linewidth]{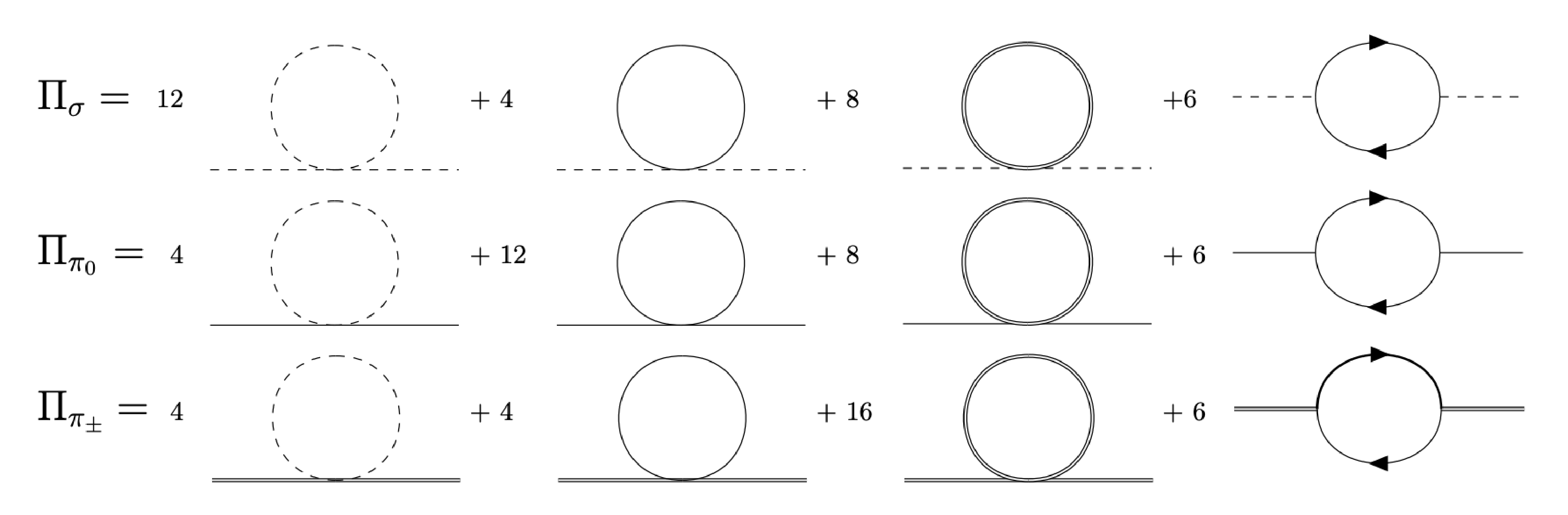}
    \caption{One-loop self-energy diagrams for the bosonic degrees of freedom. Dashed lines represent the sigma meson; single and continuous lines correspond to neutral pions; double and continuous lines denote charged pions; and continuous lines with arrows represent quark fields. The coefficients multiplying each diagram are the corresponding combinatorial factors.}
    \label{diagramsselfenergy}
\end{figure}
}
The explicit expressions for the boson self-energies, obtained by computing the derivatives in Eqs.~(\ref{self-energy-potentialbosons}) and~(\ref{self-energy-potentialfermions}), are given by
\begin{align}
    \Pi_{\sigma}=& \frac{3\lambda}{2}\left ( \frac{ T^2}{6} -\frac{ T\sqrt{m_{\sigma}^2}}{2\pi}-\frac{ m_{\sigma}^2}{8\pi^2}\left ( 1-2\gamma_E-\ln \left ( \frac{m_{\sigma}^2}{(4\pi T)^2}\right )\right )\right )+\lambda\left (\frac{ T^2}{6} -\frac{T \sqrt{m_{\pi}^2}}{2\pi}-\frac{m_{\pi}^2}{8\pi^2}\left ( 1-2\gamma_E-\ln \left ( \frac{m_{\pi}^2}{(4\pi T)^2}\right )\right )\right ) \nonumber \\
    &+\frac{\lambda|eB|}{2\pi^2} \ln \left ( \frac{\mu^2}{m_{\pi}^2+|eB|}\right )+\frac{\lambda |eB|}{2\pi^2}\sum_{n=1}^{\infty} K_0 \left ( \frac{n\sqrt{m_{\pi}^2+|eB|}}{T}\right )-6\left ( \frac{4g^2|qB|}{\pi^2} \sum_{n=1}^{\infty}(-1)^n K_0\left(\frac{ n  g f_\pi}{T}\right)  \right ),
    \label{sigma-self-energy}
\end{align}
\begin{align}
    \Pi_{\pi_0}=&\lambda \left ( \frac{T^2}{6} -\frac{T \sqrt{m_{\sigma}^2} }{2\pi}-\frac{m_{\sigma}^2}{8\pi^2}\left ( 1-2\gamma_E-\ln \left ( \frac{m_{\sigma}^2}{(4\pi T)^2}\right )\right )\right )+\frac{3\lambda}{2}\left (\frac{ T^2}{6} -\frac{T \sqrt{m_{\pi}^2}}{2\pi}-\frac{m_{\pi}^2}{8\pi^2}\left ( 1-2\gamma_E-\ln \left ( \frac{m_{\pi}^2}{(4\pi T)^2}\right )\right )\right ) \nonumber \\
    &+\frac{\lambda|eB|}{2\pi^2} \ln \left ( \frac{\mu^2}{m_{\pi}^2+|eB|}\right )+\frac{\lambda |eB|}{2\pi^2}\sum_{n=1}^{\infty} K_0 \left ( \frac{n\sqrt{m_{\pi}^2+|eB|}}{T}\right )-6\left ( \frac{4g^2|qB|}{\pi^2} \sum_{n=1}^{\infty}(-1)^n K_0\left(\frac{ n  g f_\pi}{T}\right)  \right ),
    \label{neutralpion-self-energy}
\end{align}
\begin{align}
    \Pi_{\pi_\pm}=& \lambda\left( \frac{ T^2}{6} -\frac{T\sqrt{m_{\sigma}^2} }{2\pi}-\frac{m_{\sigma}^2}{8\pi^2}\left( 1-2\gamma_E-\ln \left( \frac{m_{\sigma}^2}{(4\pi T)^2}\right)\right)\right)+\lambda\left(\frac{ T^2}{6} -\frac{T \sqrt{m_{\pi}^2}}{2\pi} -\frac{m_{\pi}^2}{8\pi^2}\left( 1-2\gamma_E-\ln \left( \frac{m_{\pi}^2}{(4\pi T)^2}\right)\right)\right) \nonumber \\
    &+\frac{\lambda|eB|}{\pi^2} \ln \left( \frac{\mu^2}{m_{\pi}^2+|eB|}\right)+\frac{\lambda |eB|}{\pi^2}\sum_{n=1}^{\infty} K_0 \left( \frac{n\sqrt{m_{\pi}^2+|eB|}}{T}\right),
    \label{chargedpion-self-energy}
\end{align}
\end{widetext}
where $K_0(x)$ is the modified Bessel function of the second kind. With the explicit expressions for the boson self-energies at one-loop order, all necessary components for constructing the effective potential beyond the mean-field approximation are now in place. These self-energies incorporate thermal and magnetic effects, and are essential for accurately describing collective phenomena such as screening effects. The resulting effective potential, which includes contributions from the tree-level, one-loop, and ring-diagram, provides a robust framework for analyzing the thermodynamics of the system. In the next subsection, we assess the validity of the approximations employed throughout the computation, before proceeding to explore the implications of our results for the restoration and breaking of chiral symmetry.

\subsection{Validity of approximations}

The one-loop contributions to the effective potential from neutral and charged pions, the sigma meson, and the quark fields were computed using two approaches: the high-temperature expansion and full numerical evaluation of the matter parts. To assess the range of validity of the high-temperature approximation, we present a comparative analysis of both methods for each sector of the theory.

Figure~\ref{validityneutralbosons} shows the comparison between the high-temperature approximation, $V^{1,0}_{b,HT}$, and the numerical evaluation $V^{1,0}_{b,N}$, for the one-loop contribution of neutral bosons to the effective potential, as a function of $v$. In panel~\ref{validityneutralbosons}$(a)$, corresponding to $T=0.1$ GeV, we observe that the region of agreement between both methods is extremely limited. As the temperature increases to $T=0.6$ GeV, shown in panel~\ref{validityneutralbosons}$(b)$, the domain of validity improves slightly, but deviations become more pronounced as $v$ increases. For $T=1.1$ GeV, panel~\ref{validityneutralbosons}$(c)$ shows that both approaches agree reasonably well, but only within a narrow window. These results indicate that the high-temperature approximation for neutral bosons cannot be applied indiscriminately, especially if one aims to explore a broad region of the effective potential. Therefore, in this case, we rely exclusively on the numerical result.

In contrast, Fig.~\ref{validitychargedbosons} displays the comparison between the high-temperature expression $V^{1,B}_{b,HT}$ and the numerical result $V^{1,B}_{b,N}$ for the charged boson contribution, again as a function of  $v$
for the same set of temperatures: $0.1$, $0.6$ and $1.1$ GeV. In all cases, we find excellent agreement across the entire range of $v$, demonstrating that both approaches can be used interchangeably in this sector.

Finally, Fig.~\ref{validityfermions} presents the analogous comparison for the fermionic contribution, between $V^1_{f,HT}$ and $V^1_{f,N}$. Once again, we observe excellent agreement at both low and high temperatures. Thus, for the fermionic sector as well, either approximation is equally valid.

These observations allow us to conclude that, while the high-temperature approximation is reliable for charged bosons and fermions, it is inadequate for the neutral boson contribution unless the temperature is sufficiently high and the region of interest in $v$ is small. Consequently, we retain the numerical result for neutral bosons in all subsequent analyses.

\begin{widetext}

\center{\begin{figure}[H]
\centering
\begin{subfigure}{0.325\textwidth}
\includegraphics[height=3.4cm]{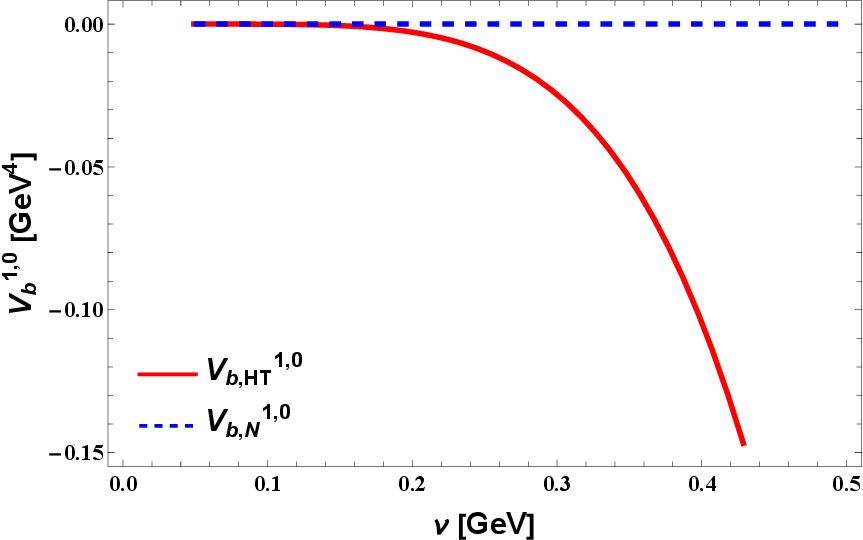}
\caption{$T=$0.1 GeV}
\end{subfigure}
\begin{subfigure}{0.325\textwidth}
\includegraphics[height=3.4cm]{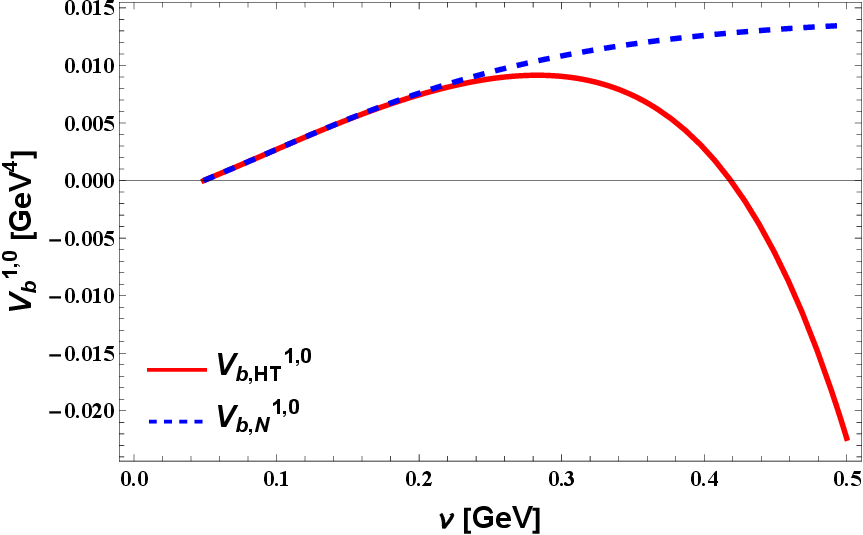}
\caption{$T=$0.6 GeV}
\end{subfigure}
\begin{subfigure}{0.325\textwidth}
\includegraphics[height=3.4cm]{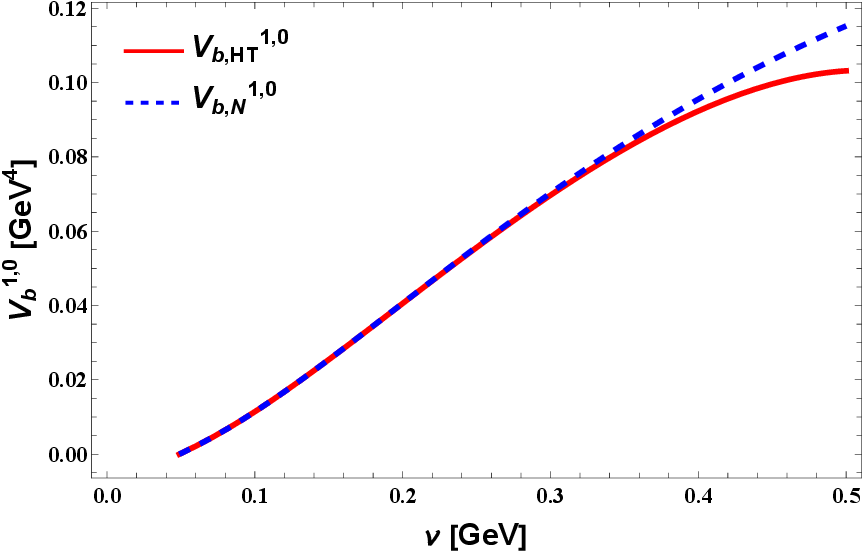}
\caption{$T=$1.1 GeV}
\end{subfigure}
\caption{Purely thermal part of the bosonic one-loop contribution to the effective potential as a function of the vacuum expectation value $v$. The dashed blue line corresponds to the numerical solution, while the solid red line represents the high-temperature approximation. Panel $(a)$ shows the result for a low temperature, $T=0.1$ GeV; panel $(b)$ illustrates the validity region for an intermediate temperature, $T=0.6$ GeV; and panel $(c)$ displays the result at high temperature, $T=1.1$ GeV. The parameters used are $\lambda=13.32$, $g=2.58$ and $a=0.309$ GeV,  obtained from the relations in Eq.~(\ref{fixedparameters}).}
\label{validityneutralbosons}
\end{figure}}

\center{\begin{figure}[H]
\centering
\begin{subfigure}{0.325\textwidth}
\includegraphics[height=3.4cm]{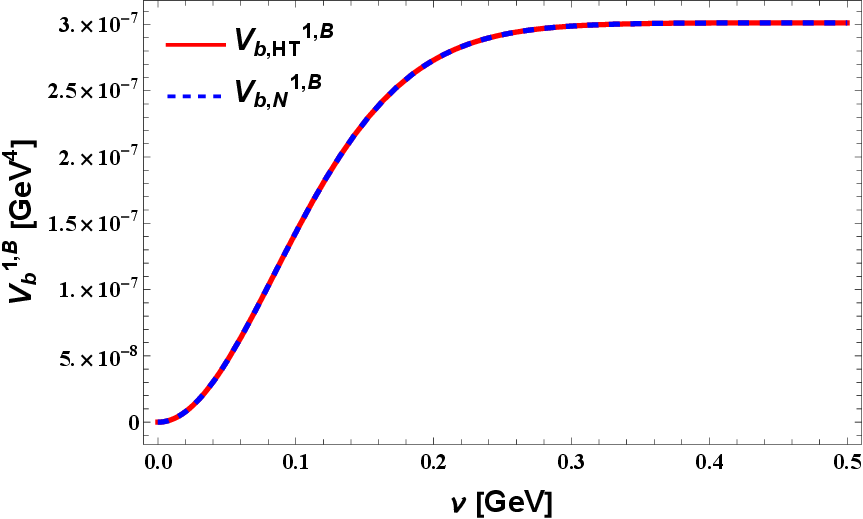}
\caption{$T=$0.1 GeV}
\end{subfigure}
\begin{subfigure}{0.325\textwidth}
\includegraphics[height=3.4cm]{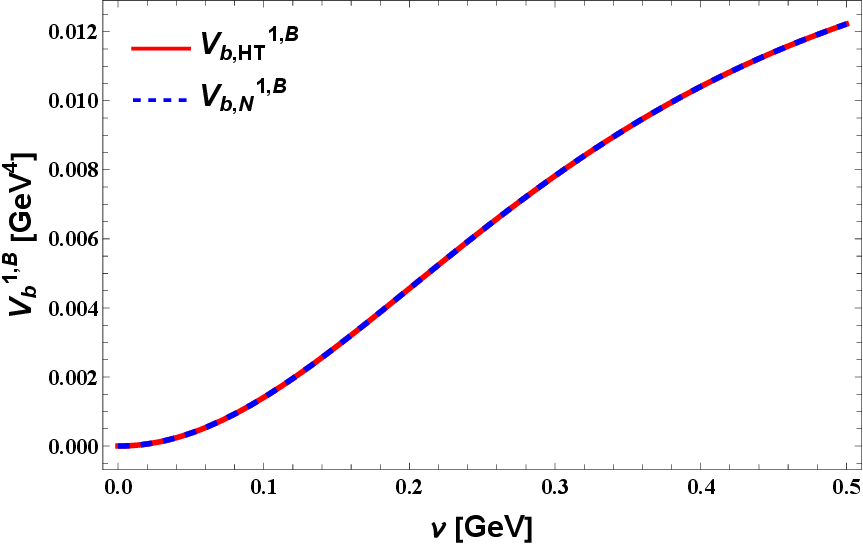}
\caption{$T=$0.6 GeV}
\end{subfigure}
\begin{subfigure}{0.325\textwidth}
\includegraphics[height=3.4cm]{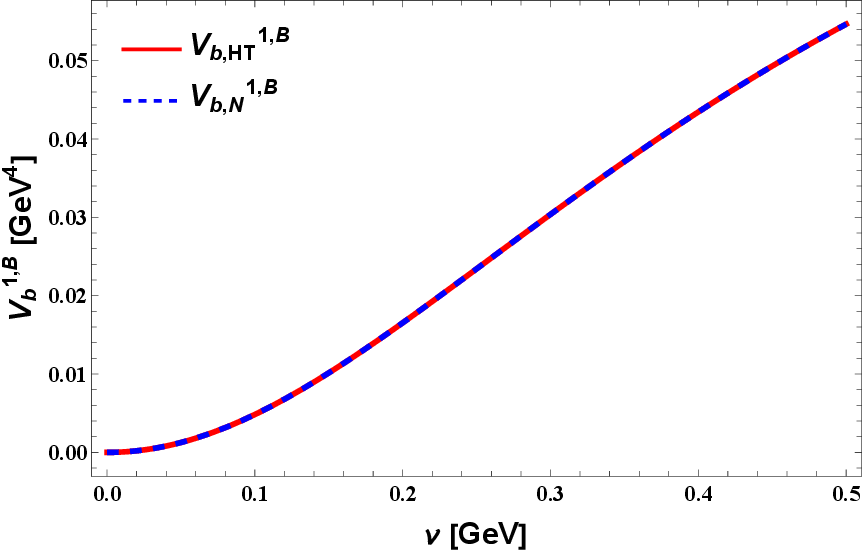}
\caption{$T=$1.1 GeV}
\end{subfigure}
\caption{Purely thermomagnetic part of the bosonic one-loop contribution to the effective potential as a function of the vacuum expectation value $v$. The dashed blue line corresponds to the numerical solution, while the solid red line represents the high-temperature approximation. Panel $(a)$ shows the result for a low temperature, $T=0.1$ GeV; panel $(b)$ illustrates the validity region for an intermediate temperature, $T=0.6$ GeV; and panel $(c)$ displays the result at high temperature, $T=1.1$ GeV. The parameters used are $\lambda=13.32$, $g=2.58$ and $a=0.309$ GeV,  obtained from the relations in Eq.~(\ref{fixedparameters}).}
\label{validitychargedbosons}
\end{figure}}

\center{\begin{figure}[H]
\centering
\begin{subfigure}{0.325\textwidth}
\includegraphics[height=3.4cm]{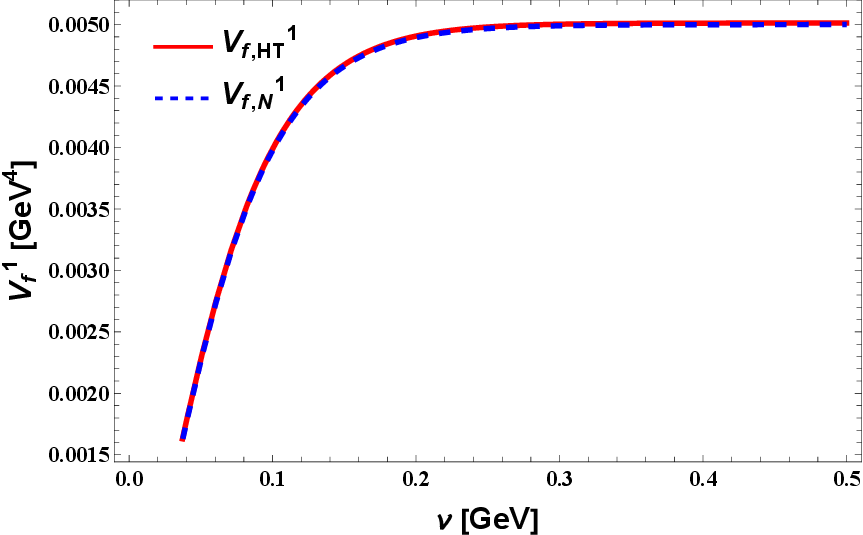}
\caption{$T=$0.1 GeV}
\end{subfigure}
\begin{subfigure}{0.325\textwidth}
\includegraphics[height=3.4cm]{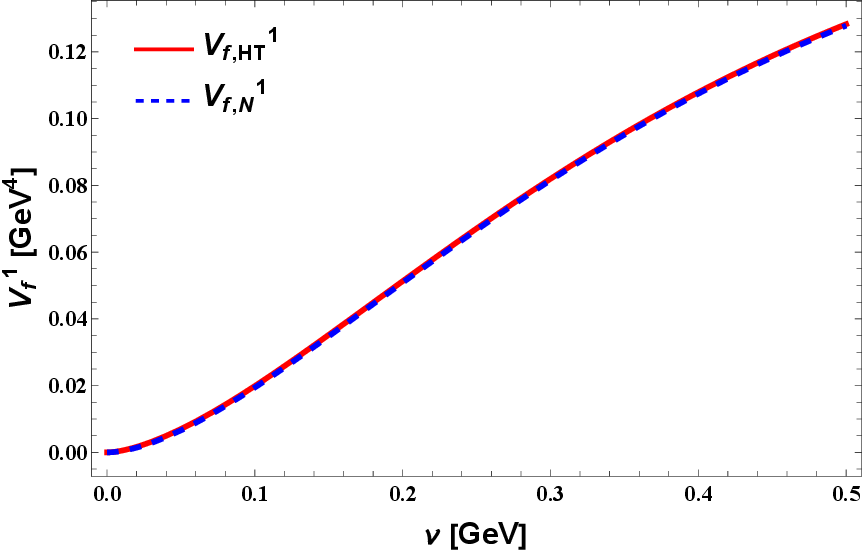}
\caption{$T=$0.6 GeV}
\end{subfigure}
\begin{subfigure}{0.325\textwidth}
\includegraphics[height=3.4cm]{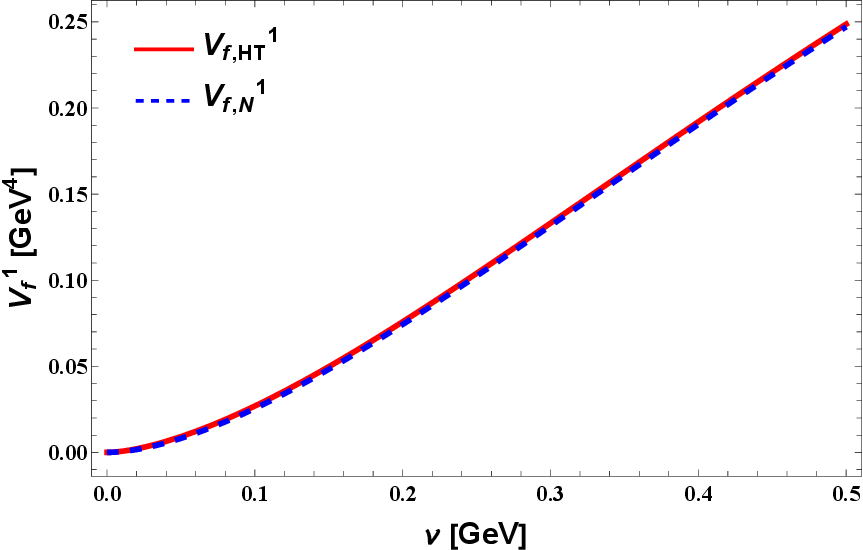}
\caption{$T=$1.1 GeV}
\end{subfigure}
\caption{Purely thermomagnetic part of the fermionic one-loop contribution to the effective potential as a function of the vacuum expectation value $v$. The dashed blue line corresponds to the numerical solution, while the solid red line represents the high-temperature approximation. Panel $(a)$ shows the result for a low temperature, $T=0.1$ GeV; panel $(b)$ illustrates the validity region for an intermediate temperature, $T=0.6$ GeV; and panel $(c)$ displays the result at high temperature, $T=1.1$ GeV. The parameters used are $\lambda=13.32$, $g=2.58$ and $a=0.309$ GeV,  obtained from the relations in Eq.~(\ref{fixedparameters}).}
\label{validityfermions}
\end{figure}}
\end{widetext}

\subsection{Preliminary results of the phase transition}

\begin{figure}[b]
    \centering
    \includegraphics[scale=0.5]{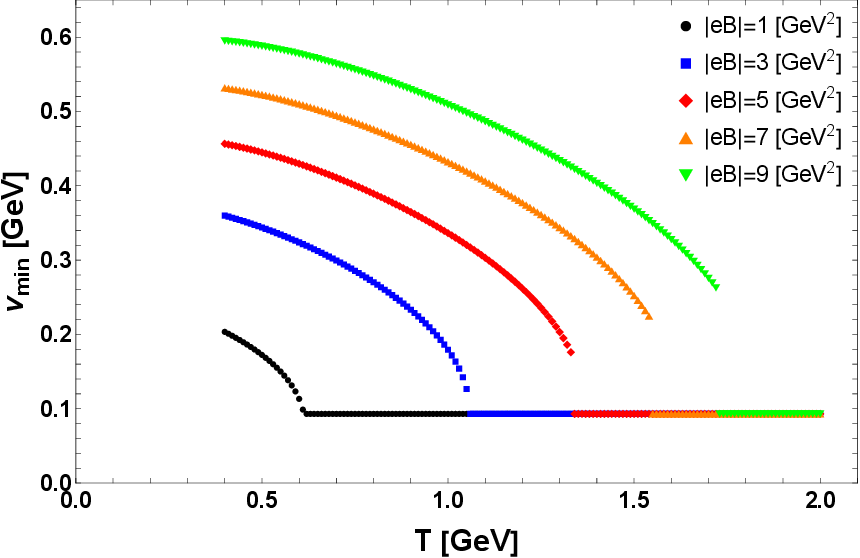}
    \caption{Minimum of the effective potential Eq.~(\ref{fulleffectivepotentialN}) as a function of the temperature, for five different values of the magnetic field $|eB|= 1- 9 \ \text{GeV}^2$. The parameters used are $\lambda=13.32$, $g=2.58$ and $a=0.309$ GeV,  obtained from the relations in Eq.~(\ref{fixedparameters}).}
    \label{vevMC}
\end{figure}

The effective potential up to ring diagrams has already been obtained under the lowest Landau level approximation and is valid for arbitrary temperature, as discussed in the previous subsection. Using Eqs.~(\ref{fulleffectivepotentialN}), (\ref{sigma-self-energy}), (\ref{neutralpion-self-energy}), and~(\ref{chargedpion-self-energy}), we are able to determine the (pseudo)critical temperature associated with the restoration of chiral symmetry. This is achieved by studying the behavior of the minimum $v_{\text{min}}$ of the effective potential $V^{\text{eff}}$ as the temperature $T$ varies, while keeping $|eB|$ fixed. The restoration of symmetry can occur either via a smooth crossover transition or through a first-order phase transition.

In the case of a crossover, $v_{\text{min}}$ evolves continuously and differentiably with respect to the thermodynamic variable $T$, and no derivative of $v_{\text{min}}(T)$ diverges. Nevertheless, the transition can still be identified through a change in the curvature of the function: the point where $v_{\text{min}}(T)$ changes from concave to convex (or vice versa). This inflection indicates a region of rapid variation in the order parameter, which is characteristic of a crossover transition. In contrast, during a first-order phase transition, $v_{\text{min}}$ does not evolve continuously with temperature. Instead, the transition is signaled by a discontinuous jump in $v_{\text{min}}(T)$, which defines the critical temperature.

Hence, in Fig.~\ref{vevMC}, we show the behavior of $v_{\text{min}}$ as a function of temperature for five different values of the magnetic field strength. We observe that for $|eB| = 1$ and $3\ \text{GeV}^2$, $v_{\text{min}}$ remains continuous and smooth. In contrast, for higher values of the magnetic field, namely $|eB| = 5$, $7$, and $9\ \text{GeV}^2$, the minimum of the effective potential exhibits a discontinuity as $T$ varies. Moreover, we observe that the curves $v_{\text{min}}(T)$ corresponding to increasing values of $|eB|$ lie on top of each other without any crossing. From these features in Fig.~\ref{vevMC}, we can identify two main characteristics of the phase transition. First, a crossover is obtained for small values of $|eB|$ (within the validity range of the LLL approximation), while a first-order phase transition emerges for larger magnetic field strengths. Second, the (pseudo)critical temperature increases monotonically with $|eB|$\footnote{Although naively the transition for $eB=1 \ \text{GeV}^2$ may look second order, a careful look indicates that the first derivative of the potential with respect to the condensate is also continuous and that the condensate never vanishes completely, which is a characteristic of a crossover. This is expected since we have introduced a term that induces explicit symmetry breaking in Eq.~(10) parametrized by the constant $h$.}.

\begin{figure}
\centering
    \includegraphics[width=0.8\linewidth]{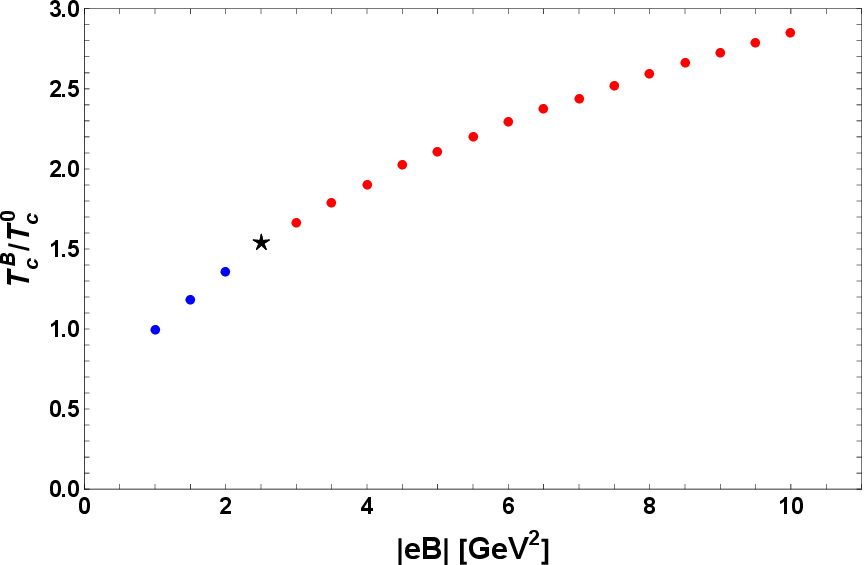}
    \caption{Phase diagram in the $T-|eB|$ plane. The vertical axis shows the (pseudo)critical temperature normalized to its value at $|eB|=0$. Blue dots indicate the crossover region, red dots correspond to the first order phase transition, and the black star marks the location of the critical end point (CEP). The parameters used are $\lambda=13.32$, $g=2.58$ and $a=0.309$ GeV,  obtained from the relations in Eq.~(\ref{fixedparameters}).}
    \label{PDMC}
\end{figure}

The information presented in Fig.~\ref{vevMC} is summarized in Fig.~\ref{PDMC}, where we show the phase diagram in the $T–|eB|$ plane. We observe that the (pseudo)critical temperature $T_c$ increases monotonically with the magnetic field strength $|eB|$, and that a critical end point (CEP) emerges around $|eB| = 3\ \text{GeV}^2$. This latter feature is consistent with the findings of Ref.~\cite{DElia:2021yvk}. However, the overall increasing behavior of $T_c$ with $|eB|$ is in contrast with the results reported in Ref.~\cite{Bali:2011qj}, where an inverse magnetic catalysis (IMC) effect is observed. In our case, we find magnetic catalysis (MC) instead.

\section{\label{sec4} Self-consistent masses}

\begin{figure}[t]
    \centering
    \includegraphics[scale=0.5]{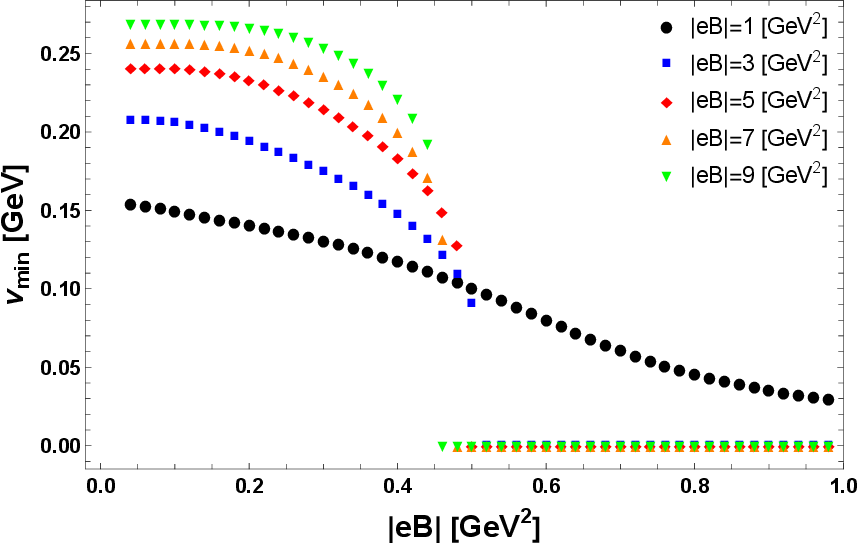}
    \caption{Minimum of the effective potential with self-consistent masses, as a function of the temperature, for five different values of the magnetic field $|eB|= 1- 9 \ \text{GeV}^2$. The parameters used are $\lambda=13.32$, $g=0.36$ and $a=0.309$ GeV.}
    \label{vevIMC}
\end{figure}

As shown in the previous section, the use of tree-level thermal masses, Eq.~(\ref{masses}), leads to magnetic catalysis for all ranges of temperature, in disagreement with LQCD results~\cite{DElia:2021yvk}. In this section, we improve upon that treatment by computing the bosonic masses self-consistently. To this end, we solve the following equations for the bosonic fields
\begin{align}
    M_{\sigma}^2(T,B)=&3 \lambda v^2-a^2+\frac{3}{2}\Pi_{b}^{0} \left ( M_\sigma^2,T\right )+\Pi_{b}^{0}\left ( 
    M_{\pi_0}^2,T\right )\nonumber \\
    &+2\Pi_{b}^B\left ( 
    M_{\pi_{\pm}}^2,T,B\right )
    +6\Pi_f(M_f,T,B),
 \label{Msigma}
\end{align}
\begin{align}
    M_{\pi_{0}}^2(T,B)=& \lambda v^2-a^2+\Pi_b^0\left ( 
 M_\sigma^2,T\right )+\frac{3}{2}\Pi_b^0\left ( 
M_{\pi_0}^2,T\right )\nonumber \\
&+2\Pi_{b}^B\left ( 
M_{\pi_{\pm}}^2,T,B\right )+6\Pi_f(M_f,T,B),
 \label{Mpi0}
\end{align}
\begin{align}
    M_{\pi_{\pm}}^2(T,B)=& \lambda v^2-a^2+\Pi_b^0\left ( 
 M_\sigma^2,T\right )+\Pi_b^0\left ( 
M_{\pi_0}^2,T\right )\nonumber \\
&+6\Pi_b^B\left ( 
 M_{\pi_{\pm}}^2,T,B\right ),
 \label{Mpic}
\end{align}
where the self-energy expressions are the following. For the neutral bosons
\begin{align}
    \Pi_b^0=&\dfrac{\lambda}{\pi^2}\int_{-\infty}^{\infty} dk\, \frac{k^2}{\sqrt{k^2+M_b^2}}\frac{1}{e^{\sqrt{k^2+M_b^2}/T}-1},
\end{align}
for charged bosons
\begin{align}
\Pi_b^B=&\frac{\lambda|eB|}{4\pi^2} \ln \left( \frac{\mu^2}{M_{b}^2+|eB|}\right)\nonumber \\
&+\frac{\lambda |eB|}{2\pi^2}\sum_{n=1}^{\infty} K_0 \left( \frac{n\sqrt{M_{b}^2+|eB|}}{T}\right),
\end{align}
and for fermions
\begin{align}
    \Pi_f=&-\frac{4g^2|qB|}{\pi^2} \sum_{n=1}^{\infty}(-1)^n K_0\left(\frac{ n  M_f}{T}\right).
\end{align}
The unknowns in Eqs.~(\ref{Msigma})-(\ref{Mpic}) are $M_\sigma$, $M_{\pi_0}$ and $M_{\pi_\pm}$, which correspond to the effective boson masses incorporating the influence of the thermal bath and external magnetic field. These equations are solved numerically. Once the solutions are obtained, we implement the substitutions $m_\sigma \rightarrow M_\sigma$, $m_{\pi_0}\rightarrow M_{\pi_0}$ and  $m_{\pi_\pm}\rightarrow M_{\pi_\pm}$ in the effective potential, Eq.~(\ref{fulleffectivepotentialN}). Note that we do not compute the fermion mass self-consistently in this work, as such calculation lies beyond the present scope and is left for future investigation. Nevertheless, to account for thermomagnetic effects on the fermion mass, analogous to the self-consistent treatment, we introduce a phenomenological ansatz of the form $M_f = g v+ \alpha + \eta |eB|$, with $g=0.36$, $\alpha=0.233$ GeV and $\eta=-0.014$ GeV$^{-1}$. Accordingly, we also perform the replacement $m_f \rightarrow M_f$ in the effective potential. It is important to note that, even with the self-consistent treatment of the boson masses and the phenomenological modification of the fermion mass, all calculations are still performed within the validity regime of the lowest Landau level (LLL) approximation. Furthermore, an implication specific to the phenomenological ansatz for the fermion mass is the need to adjust the coupling constant $g$, which must be taken to be smaller than the value determined in vacuum. With these modified masses, we repeat the analysis to determine the phase transition line in the $T-|eB|$ plane. The results are shown in Figs.~\ref{vevIMC} and~\ref{PDIMC}. The first figure shows that, for the two smallest values of $|eB|$ considered, the transition is a crossover, while for larger values, it becomes a first-order phase transition. Additionally, we observe that curves of $v_{\text{min}}$ for different values of $|eB|$ intersect, indicating that the (pseudo)critical temperature decreases as the magnetic field increases. Consequently, in Fig.~(\ref{PDIMC}), we observe the phenomenon of inverse magnetic catalysis (IMC), characterized by a decreasing $T_c$ as a function of $|eB|$, and the emergence of a critical end point (CEP) associated with the change in the nature of the phase transition as the magnetic field grows.

\section{\label{sec5} Results}

We now proceed to present and analyze the numerical results obtained from the effective potential constructed with the procedures described above.

\subsection{Analysis with tree-level thermal masses}

In this subsection, we analyze the thermodynamic behavior of the model using the effective potential constructed with tree-level field-dependent masses. The free parameters of the model were fixed in vacuum, and the potential was minimized with respect to the order parameter $v$ at fixed values of temperature and magnetic field.

Figure~\ref{vevMC} shows the behavior of the vacuum expectation value $v_{\text{min}}$ as a function of temperature for several values of the magnetic field strength. From this figure, we observe that the transition becomes sharper as $|eB|$ increases, and a discontinuity emerges at large magnetic fields. This is a clear indication that the crossover at low $|eB|$ evolves into a first-order phase transition. In Fig.~\ref{PDMC}, we present the phase diagram in the $T-|eB|$ plane. We find that the pseudocritical temperature increases monotonically with the magnetic field strength, a behavior commonly referred to as magnetic catalysis. Moreover, a critical endpoint appears at a finite magnetic field value, signaling the end of the crossover line and the onset of a first-order transition. 

\begin{figure}
    \centering
    \includegraphics[scale=0.5]{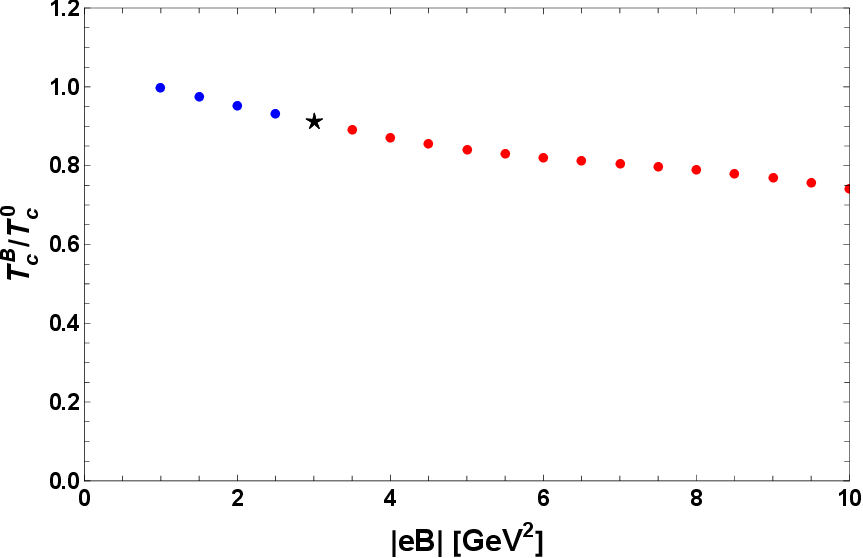}
    \caption{Improved phase diagram in the $T-|eB|$ plane, incorporating the self-consistently computed masses. The vertical axis shows the (pseudo)critical temperature normalized to its value at $|eB|=0$. Blue dots indicate the crossover region, red dots correspond to the first order phase transition, and the black star marks the location of the critical end point (CEP). The parameters used are $\lambda=13.32$, $g=0.36$ and $a=0.309$ GeV.}
    \label{PDIMC}
\end{figure}

The emergence of a critical endpoint is a notable and nontrivial feature, which has also been reported in LQCD studies. However, the magnetic catalysis behavior obtained here is in contradiction with the LQCD results, which indicate that the pseudocritical temperature decreases with increasing $|eB|$, a phenomenon known as inverse magnetic catalysis.

This discrepancy highlights a limitation of the tree-level mass treatment and suggests that the inclusion of dynamical collective effects is required to capture the correct thermomagnetic behavior. We now proceed to incorporate these effects through a self-consistent computation of the boson masses.

\subsection{Results with self-consistent boson masses}

Using the self-consistent masses obtained in Sec.~\ref{sec4}), we recompute the effective potential and analyze the chiral phase transition under the same external conditions.

Figure~\ref{vevIMC} displays the behavior of the order parameter $v_{\text{min}}$ as a function of temperature for various values of the magnetic field. In contrast to the tree-level case, we now observe that the pseudocritical temperature decreases with increasing $|eB|$, in agreement with the inverse magnetic catalysis phenomenon reported in LQCD. The smooth behavior at low fields and the appearance of a discontinuity at high fields remain present, indicating that the critical endpoint persists in the improved analysis.

The resulting phase diagram in the $T-|eB|$ plane is shown in Fig.~\ref{PDIMC}. It reveals that the transition line bends downward as the magnetic field increases, in qualitative agreement with lattice QCD results. Furthermore, the critical endpoint is still present, located at a slightly larger magnetic field compared to the case without self-consistent masses.

These results confirm that the self-consistent treatment of the boson masses captures essential collective effects induced by temperature and magnetic field. Remarkably, this improvement allows us to reproduce both the inverse magnetic catalysis and the existence of a critical point without the need to introduce ad hoc effective couplings. This highlights the predictive power of the model when applied beyond the mean-field approximation.

\section{\label{sec6} Conclusions}

In this work, we have revisited the phase structure of strongly interacting matter under the influence of an external magnetic field within the framework of the linear sigma model coupled to quarks, going beyond the traditional mean-field approximation. A key novelty of our approach is the self-consistent computation of the bosonic thermal masses in the presence of a magnetic field. To the best of our knowledge, this is the first time such a treatment has been implemented in the literature using this model in a magnetic background. The self-consistent procedure involves solving a coupled set of gap equations that incorporate both thermal and magnetic effects on the boson propagators and implementing the resulting masses back into the effective potential. For the fermion sector, although a full self-consistent calculation lies beyond the scope of the present work, we introduced a phenomenological magnetic ansatz to mimic the qualitative behavior expected from non-perturbative effects.

Our results demonstrate that this improved treatment leads to a significant shift in the qualitative behavior of the phase transition. While the use of tree-level thermal masses leads to magnetic catalysis (MC), the inclusion of self-consistent boson masses and a magnetic fermion mass yields a clear manifestation of inverse magnetic catalysis (IMC), in qualitative agreement with lattice QCD results~\cite{DElia:2021yvk}. Furthermore, we observe the emergence of a critical end point (CEP) in the phase diagram, located at intermediate values of the magnetic field strength, and find that the transition changes from a smooth crossover to a first-order phase transition as the field increases. These features emerge naturally in our framework without invoking ad hoc magnetic-field-dependent couplings, and arise purely from the internal consistency of the resummed effective potential.

This work opens a new direction for systematically improving effective models of QCD under extreme conditions. The framework developed here can be extended to include finite baryon chemical potential, rotation, or even chiral imbalance, and revisiting many existing results in the literature with self-consistent mass treatments could provide more accurate and physically meaningful insights. Our approach provides a robust and versatile tool for exploring the rich structure of QCD matter in environments such as heavy-ion collisions and the interior of neutron stars, where magnetic fields and thermal effects play a fundamental role.

\begin{acknowledgments}
Support for this work was received in part by the Secretaria de Ciencia, Humanidades, Tecnología e Innovación Grants No. CF-2023-G-433 and No. CBF-2025-G-1718. LAH acknowledges support from the DAI UAM PIPAIR 2024 project under Grant No. TR2024-800-00744. AM acknowledges support from grant 2023/08826-7, from the São Paulo Research Foundation (FAPESP), and GF acknowledges the financial support of a fellowship granted by Secretaria de Ciencia, Humanidades, Tecnología e Innovación as part of the Sistema Nacional de Posgrados. 
\end{acknowledgments}

\appendix


\section{High temperature approximation for the charged boson matter term at one-loop order. \label{appendix2}}

The matter contribution to the effective potential from the charged boson sector at one-loop order can be expressed as an infinite series. Starting from Eq.~(\ref{V1BafterMatsubara}), where the matter piece can be written as
\begin{equation}
   V_{b,N}^{1,B}=\frac{T|eB|}{2\pi}\int \frac{dk_3}{2\pi}\ln \left(1-e^{-\sqrt{k_3^2+m_b^2+|eB|}/T} \right), 
\end{equation}
whose integrand can be expanded using the identity
\begin{equation}
    \ln(1+x)=\sum_{n=1}^\infty \frac{(-1)^{n-1}}{n}x^n,
    \label{expansionln}
\end{equation}
we obtain for $V_{b,N}^{1,B}$
\begin{equation}
    V_{b,N}^{1,B}=-\frac{T|eB|}{2\pi}\int\frac{dk_3}{2\pi}\sum_{n=1}^\infty \frac{e^{-\sqrt{k_3^2+m_b^2+|eB|}n/T}}{n}.
\end{equation}
Introducing the change of variable $z=\sqrt{k_3^2+m_b^2+|eB|}$, the expression becomes
\begin{equation}
    V_{b,N}^{1,B}=-\frac{T|eB|}{2\pi^2}\sum_{n=1}^\infty \int_{\sqrt{m_b^2+|eB|}}^\infty \frac{dz}{n} \frac{z \ e^{-nz/T}}{\sqrt{z^2-m_b^2-|eB|}}.
\end{equation}
Evaluating the integral over $z$, we obtain the compact result
\begin{align}
    V_{b,N}^{1,B}&=-\frac{T|eB|}{2\pi^2}\sqrt{m_b^2+|eB|}\nonumber \\
    &\times \sum_{n=1}^\infty \frac{K_1(n\sqrt{m_b^2+|eB|}/T)}{n},
\end{align}
where $K_1$ denotes the modified Bessel function of the second kind.

\section{High temperature approximation for the fermion matter term at one-loop order. \label{appendix3}}

The matter contribution from the fermionic sector at one-loop order can also be expressed as an infinite series. To derive this result, we follow the same procedure used for the charged boson case, described in the previous Appendix. Starting from Eq.~(\ref{Vf1afterMatsubarasum}), where the matter piece can be expressed as
\begin{equation}
    V_{f,HT}^1=-\frac{2N_c|qB|T}{\pi}\int \frac{dk_3}{2\pi}\ln\left(1+e^{-\sqrt{k_3^2+m_f^2}/T}\right),
\end{equation}
and using the expansion of the logarithm function shown in Eq.~(\ref{expansionln}), we obtain
\begin{equation}
    V_{f,HT}^1=-\frac{2N_c|qB|T}{\pi}\int\frac{dk_3}{2\pi}\sum_{n=1}^\infty \frac{(-1)^n}{n}e^{-\sqrt{k_3^2+m_f^2}n/T}.
\end{equation}
Introducing the change of variable $z=\sqrt{k_3^2+m_f^2}$, this becomes
\begin{equation}
    V_{f,HT}^1=-\frac{2N_c|qB|T}{\pi^2}\sum_{n=1}^\infty\int_{m_f}^\infty dz \frac{z \ e^{-nz/T}}{\sqrt{z^2-m_f^2}}.
\end{equation}
Evaluating the integral over $z$, we obtain the final result
\begin{align}
    V_{f,HT}^1&=-\frac{2N_c|qB|T}{\pi^2}m_f\nonumber \\
    &\times \sum_{n=1}^\infty\frac{(-1)^n}{n}K_1\left(\frac{n \ m_f}{T}\right),
\end{align}
where $K_1$ is the same modified Bessel function introduced in the previous Appendix.

\bibliography{mybibliography}

\end{document}